\documentclass[aps,prl,twocolumn,showpacs,preprintnumbers,superscriptaddress,amsmath]{revtex4}
\usepackage{graphicx}
\usepackage{dcolumn}
\usepackage{bm}
\begin{document}
\title{Coupling of light from an optical fiber taper into silver
nanowires}

\author{Chun-Hua Dong}
\author{Xi-Feng Ren\footnote{renxf@ustc.edu.cn}}
\author{Rui Yang}
\address{Key Laboratory of Quantum Information, University of Science
and Technology of China, Hefei 230026, People's Republic of China}
\author{Jun-Yuan Duan}
\author{Jian-Guo Guan}
\address {State Key Laboratory of Advanced Technology for Materials
Synthesis and Processing, Wuhan University of Technology, 122 Luoshi
Road, Hubei, Wuhan 430070, People's Republic of China}
\author{Guang-Can Guo}
\author{Guo-Ping Guo\footnote{gpguo@ustc.edu.cn} }
\address{Key Laboratory of Quantum Information, University of Science
and Technology of China, Hefei 230026, People's Republic of China}

\begin{abstract}
We report the coupling of photons from an optical fiber taper to
surface plasmon modes of silver nanowires. The launch of propagating
plasmons can be realized not only at ends of the nanowires, but also
at the midsection. The degree of the coupling can be controlled by
adjusting the light polarization. In addition, we present the
coupling of light into multiple nanowires from a single optical
fiber taper simultaneously. Our demonstration offers a novel method
for optimizing plasmon coupling into nanoscale metallic waveguides
and promotes the realization of highly integrated plasmonic devices.
\end{abstract}
\pacs{78.67.Lt, 73.20.Mf, 73.22.Lp}

\maketitle

With the increasing attention and progress of nanotechnology, the
dimensions of ultrafast transistors are on the order of 50 nm. The
imperative problem now is carrying digital information from one end
to the other end of a microprocessor if we want to increase the
speed of microprocessors. Optical interconnects such as fiber optic
cables can carry digital data with a capacity 1000 times more than
that of electronic interconnects, while fiber optic cables are
larger due to the optical diffraction limit. This size-compatibility
problem may be solved if the optical elements can be integrated on
chip and fabricated at nanoscale. One such proposal is surface
plasmons, which are electromagnetic waves that propagate along the
surface of a conductor\cite{ozbay}. Plasmonics, surface
plasmon-based optics, have been demonstrated and investigated
intensively in nanoscale metallic hole
arrays\cite{Ebbesen98,Moreno,Alt}, metallic
waveguides\cite{Pile,Bozhevo,Lamp}, and metallic
nanowires\cite{Dickson,Graff,Ditlbacher,Sanders,Knight,Pyayt} in
recent years. Among the different kinds of plasmonic waveguides,
sliver nanowires have some unique properties that make them
particularly attractive, such as low propagating loss due to their
smooth surface and scattering of plasmons to photons only at their
sharp ends. Since the momentums of the photons and plasmons are
different, it is a challenge to couple light into plasmon waveguides
efficiently. The general methods for plasmon excitation include
prism coupling and focusing of light onto one end of the nanowire
with a microscope objective. Nanoparticle antenna-based approach is
also proved to be an efficient way for optimizing plasmon coupling
into nanowires\cite{Knight}, which allows for direct coupling into
straight, continuous nanowires by using a nanoparticle as an
antenna. Recently, a single polymer waveguide is used to couple
light into multiple nanowires simultaneously\cite{Pyayt} as well,
aiming at providing light to a number of nanoscale devices in the
future integrated photonic circuits. Whereas due to the random
distribution of nanowires and nanoparticles, it is hard to achieve
optimum coupling efficiency for the two methods under present
technology.

Here we report a new experimental method to couple light with
plasmons in sliver nanowires by using an optical single mode fiber
taper contacting one or several nanowires. It is found that the
plasmons can be excited from the midsection of a continuous, smooth
nanowire. Using a fiber taper, we can couple light into a nanowire
from any position of it. Moreover, the fiber taper can be used to
arrange the position of the nanowires, and several nanowires can be
excited simultaneously by one fiber taper. This structure bridges
the classical optical fibers and the nanoscale plasmonic nanowires
and might be useful for coupling light to nanophotonic devices in
integrated circuits.

\begin{figure}
\includegraphics[width=6.0cm]{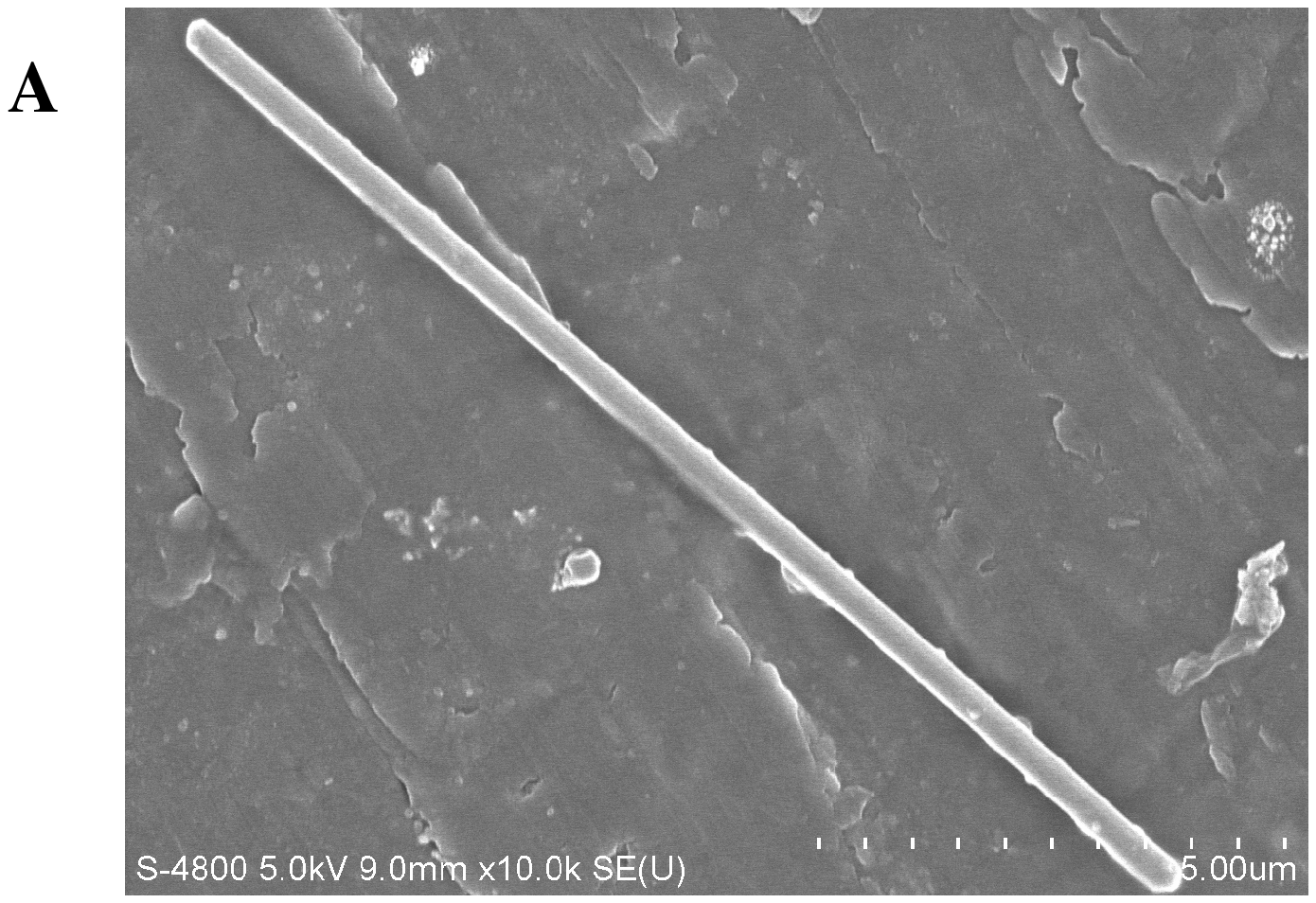}
\includegraphics[width=6.0cm]{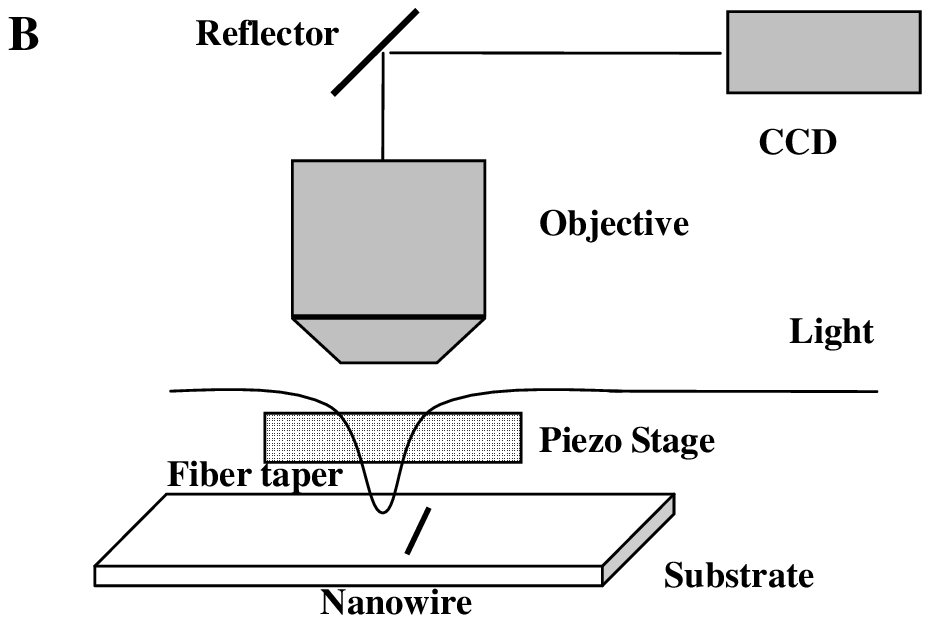}
\caption{(A)Scanning electron micrographs of a 14 $\mu m$ long
silver nanowire. Its diameter is about 300 $nm$. (B)Sketch map of
our experiment setup. Laser beam with 780 nm wavelength is coupled
into an optical fiber taper which is contacted with nanowires. The
fiber taper is mounted in a U-shaped configuration and moved by a
piezo-electric stage. Scattering light is recorded by a CCD camera
after a microscope objective.}
\end{figure}

There are a lot of methods for the controlled synthesis of silver
narowires\cite{sun,xia,huang}. Here a solvothermal process is used
to fabricate silver nanowires. In a typical synthesis procedure, 2
mmol of PVP and 1.4 mmol of AgNO$_{3}$ were successively dissolved
in 36 mL of ethylene glycol. Then 2 mL of NaCl ethylene glycol
solution (1.2 mmol/L) and 2 mL of ferric nitrate ethylene glycol
solution (15 mmol/L) were added under magnetic stirring. The mixture
was sealed in a 50 mL autoclave and heated in oven at 180 $^\circ C$
for 12 hours. Finally, the Teflon-lined autoclave was cooled
naturally to room temperature, and the final products were obtained
after centrifugation of the straw yellow suspension and washed with
deionized water and ethanol for several times (centrifugal speed is
6000 r/min). The products were preserved in ethanol. The
as-synthesized products were characterized by field emission
scanning electron microscopy (FE-SEM; Hitachi, S-4800) at an
acceleration voltage of 5.0 kV. The wires obtained here have a
diameter about 300 nm and lengths about 10 $\mu m$(Fig. 1A).

Samples used in our experiment were prepared by drop-casting a
dilute nanowires suspension on cover glass and then letting them dry
in the open air. A tapered fiber was prepared from a single mode
fiber at a wavelength of 780 nm (Newport) which was heated by a
Hydrogen microtorch and stretched to the opposite directions with
two translators\cite{cai,min}. The curvature of the taper profile
was small to realize adiabatic propagation of light through the
tapered region. In our experiment, the fiber taper reached a minimum
diameter of only about 1 $\mu m$ which had evanescent fields
outside\cite{Lou}. Laser beam with 780 nm wavelength was coupled
into the optical fiber and laser polarization was controlled by a
polarization beam splitter (PBS) followed by a half wave plate
(HWP)\cite{Konishi}. Rotating the HWP allowed us to investigate the
relationship between the coupling efficiency and the polarization of
light. The optical fiber taper was placed above and parallel to the
substrate where nanowires were dropped. It was mounted in a U-shaped
configuration and moved by a three dimensional piezo-electric stage
(Physik Instrumente Co., Ltd. NanoCube XYZ Piezo Stage), sketched in
Fig. 1B. Scattering light from the nanowire was recorded by a CCD
camera after a microscope objective.

\begin{figure}
\includegraphics[width=8.0cm]{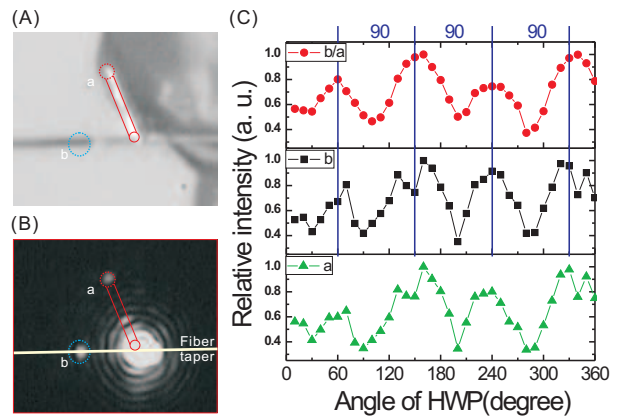}
\caption{Polarization dependence of coupling efficiency at nanowire
end. (A) Micrograph of a nanowire contacted with a fiber taper at
one end. (B) Emission can be observed from the other end. (C)
Far-field emission intensities as a function of laser polarization
angle. a is the emission intensity determined by averaging the four
brightest pixels at site "a" and b is the intensity of site "b"
which is used as a reference of background scattering since there is
a dust adhering to the fiber taper. a/b gives the relationship
between the coupling strength and the polarization of light.}
\end{figure}

To eliminate the influence of the glass surface, we put a nanowire
on the edge as shown in Figure 2A. The length of the nanowire was
about 11 $\mu m$. The fiber taper contacted the nanowire at one end
and the emission was observed from the other end clearly, which
verified that optical fiber taper could also excite surface plasmons
in metallic nanowires and couple optical information into nanoscale
devices. The coupling strength was measured by changing the
polarization of the input light. For each polarization, the emission
intensity was determined by averaging the four brightest pixels at
site "a"(see inset of Fig. 2). Intensity of site "b" was used as a
reference of background scattering since there was a dust adhering
to the fiber taper. It changed with the polarization of the input
light for the different coupling efficiencies. Fig. 2C showed the
relationship between the coupling strength and the polarization of
light. The far-field emission curve as a function of polarization
angle was approximately in accord with the theoretical
prediction(cosine or sine function)\cite{Sanders,Knight}, and the
error here might come from the strong background scattering. This
phenomenon was similar with the case that we excited surface
plasmons with focused laser spot in free space using a 100x
microscope objective.

As we know, the momentum of the propagating plasmon($k_{sp}$) is
larger than that of the incoming photon($k_{ph}$), so there needs an
additional wavevector($\Delta k$) to sustain the momentum
conservation condition. Surface plasmons in nanowires can be excited
where the symmetry is broken, for example, at the ends and sharp
bends\cite{Dickson,Graff,Ditlbacher,Sanders}, because an extra
wavevector ($\Delta k_{scatter}$) is supplied according to the
scattering mechanism in this situation. Surface plasmons can not be
excited in the midsection directly, as a result of the smooth
surface of the nanowire. Since the plasmonic waveguides may be very
long in practice like optical fiber, it will be more convenient if
we can couple light into them from the midsection. One scheme to
directly couple light into straight, continuous nanowires is using a
nanoparticle as an antenna\cite{Knight}. However this method may be
not expedient if we want to couple light at random sections of a
nanowire since the distributions of nanowires and nanoparticles can
not be controlled easily.

\begin{figure}
\includegraphics[width=8.0cm]{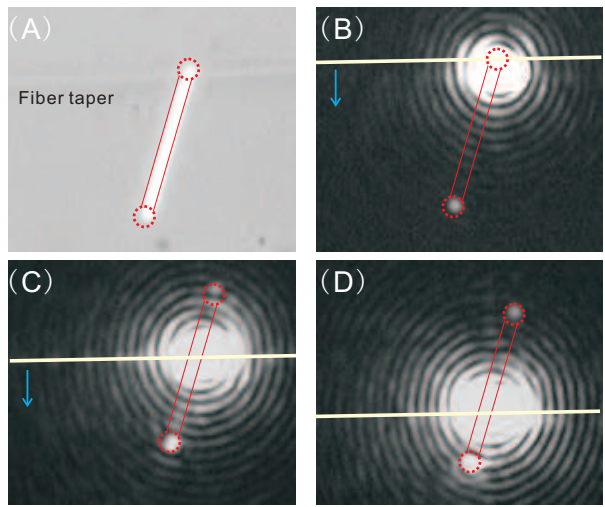}
\caption{Plasmons are observed from both ends by contacting the
fiber taper with a nanowire in the midsection. (A) Micrograph of a
nanowire contacted with a fiber taper. (B),(C),(D) The fiber taper
contact the nanowire with different sections and scattering light
from both ends are detected.}
\end{figure}
From Fig. 3, we can see that plasmons were observed from both ends
by contacting the fiber taper with a nanowire in its midsection. The
fiber taper was also moved from one end to the other end of this
nanowire slowly, and scattering light was observed as periodic glint
during this process. To testify that it was not the result of the
exceptive discontinuity of the nanowire, we focused the laser light
on the midsection using a 100X microscope objective and no plasmon
was launched. Several other nanowires were tested subsequently as
well and gave the similar phenomena. The reason for direct coupling
in midsection may be that the symmetry of the nanowire is disrupted
when the fiber taper and the nanowire are contacted with each other.
An additional momentum is offered partly by scattering on the
nanowire surface and else from the evanescent optical field of the
fiber taper. Moving the optical fiber taper by the stage randomly,
we can launch plasmons from any section of a straight nanowire.
Similar to the case of exciting plasmon from the ends, coupling
strength can be modulated by the light polarization. To check
whether the continuities of the nanowires were damaged after
contacted with fiber taper, we used the free space coupling method
and proved that the whole process of coupling light from fiber taper
to surface plasmons was safe for nanowires and exercisable in
practice. It should be noticed that the intensity of output light
from two ends changed with coupling positions. A potential
explanation is that the silver nanowire can work as an efficient
Fabry-Perot resonator, in which the scattered light intensity is
modulated as a function of coupling position with the distinct
Fabry-Perot resonator modes. Further investigation is necessary to
give a numerical analysis which is beyond this work. According to
the free coupling property of this protocol, it is especially useful
for coupling light into nanodevices which have no sharp end, such as
nanoring\cite{Mclell,wang}.

\begin{figure}
\includegraphics[width=8.0cm]{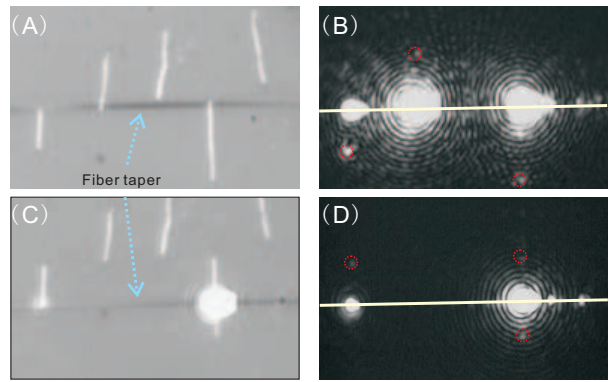}
\caption{Coupling of light into multiple nanowires from a single
optical fiber taper. (A) A micrograph with white light illumination
shows a fiber taper contacted with three nanowires simultaneously.
(B) Light scattered from the ends of the three nanowires can be
observed. (C) A fiber taper contacted with a nanowire at end and
another nanowire at midsection. (D) Dark filed picture of (C)
indicates that we can excite surface plasmons selectively from end
of a nanowire or its midsection.}
\end{figure}

Besides the benefit of coupling light from any sections of
nanowires, another advantage of the fiber taper coupling method is
that we can excite surface plasmons in many nanowires simultaneously
using a single fiber taper. In the future plasmonic circuits, we may
need to integrate many nano-waveguides to increase data transmission
rates and capacity. Obviously, the previous methods of prism
coupling and focusing with microscope objective are not convenient
and can not be applied on chips. Pyayt and his coworkers proposed to
excite plasmons in many nanowires by putting them perpendicular to a
polymer waveguide with one end located close to the light inside the
waveguide\cite{Pyayt}. In their structure, the sliver nanowires were
oriented randomly on the substrate and a series of SU-8 stripes were
covered on them as polymer waveguides. They observed that the light
coupled in the waveguide could propagate along several nanowires
simultaneously. While due to the random distribution of nanowires,
many of them did not couple light out of the waveguide. Precise
control of nanowire orientation was essential to achieve optimum
coupling efficiency. Here, we used the fiber taper to substitute for
the waveguide and discovered the similar phenomenon while the whole
process can be controlled more precisely.

We utilized a broken fiber taper to adhibit a nanowire, then moved
it to the appropriate place carefully by a nanoscale piezo stage and
put it down on the substrate. Repeating this process several times,
we got a well organized distribution of nanowires. Though some of
the nanowires might be destroyed during this operation, we can clear
the bad ones and keep the good ones. In this work, five nanowires
were placed parallel to each other on the substrate as shown in Fig.
4A. A fiber taper contacted three of them simultaneously on their
ends. We could see that light scattered from the other ends of these
three nanowires at the same time and the two un-contacted nanowires
remained dark, as shown in Fig. 4B. Likewise, we can excite surface
plasmons selectively from end of a nanowire or its midsection. This
proved that our method can be used to couple laser light to multiple
nanowires simultaneously.

In summary, we have demonstrated an original technique to couple
light into silver nanowires. The new method has two remarkable
advantages: One is that plasmons can be launched from any part of a
nanowire, and the other is that one optical fiber taper can be
applied to couple light into many nanowires simultaneously. This
method can directly combine the classical optical elements with the
nanoscale plasmonic devices, and thus may be practical for optical
input of nanoscale photonic devices in highly integrated circuits.

\begin{center}
\textbf{Acknowledgments}
\end{center}

The authors thank Prof. Younan Xia for useful discussion. This work
was funded by the National Basic Research Programme of China (Grants
No.2009CB929600 and No. 2006CB921900), the Innovation funds from
Chinese Academy of Sciences, and the National Natural Science
Foundation of China (Grants No. 10604052 and No.10874163).


\end{document}